\documentclass{ws-ijmpd}

\begin{document}

\markboth{Bianco, Massucci \& Ruffini}
{The luminosity evolution over the EQTS in GRBs prompt emission}

\title{\MakeUppercase{The luminosity evolution over the EQuiTemporal Surfaces in the prompt emission of Gamma-Ray Bursts}}

\author{\MakeUppercase{Carlo Luciano Bianco},$^{1,2}$ \MakeUppercase{Francesco Alessandro Massucci},$^{3}$ \MakeUppercase{Remo Ruffini}$^{1,2}$}

\address{
$^{1}$ICRANet, Piazzale della Repubblica 10, I-65100 Pescara, Italy. E-mail: bianco@icra.it.\\
$^{2}$Dipartimento di Fisica, Universit\`a di Roma ``La Sapienza'', Piazzale Aldo Moro 5, I-00185 Roma, Italy.\\
$^{3}$Department of Mathematics, King's College London, Strand, London WC2R2LS, United Kingdom.
}

\maketitle

\begin{history}
\received{...}
\revised{...}
\comby{...}
\end{history}

\begin{abstract}
Due to the ultrarelativistic velocity of the expanding ``fireshell'' (Lorentz gamma factor $\gamma \sim 10^2 - 10^3$), photons emitted at the same time from the fireshell surface do not reach the observer at the same arrival time. In interpreting Gamma-Ray Bursts (GRBs) it is crucial to determine the properties of the EQuiTemporal Surfaces (EQTSs): the locus of points which are source of radiation reaching the observer at the same arrival time. In the current literature this analysis is performed only in the latest phases of the afterglow. Here we study the distribution of the GRB bolometric luminosity over the EQTSs, with special attention to the prompt emission phase. We analyze as well the temporal evolution of the EQTS apparent size in the sky. We use the analytic solutions of the equations of motion of the fireshell and the corresponding analytic expressions of the EQTSs which have been presented in recent works and which are valid for both the fully radiative and the adiabatic dynamics. We find the novel result that at the beginning of the prompt emission the most luminous regions of the EQTSs are the ones closest to the line of sight. On the contrary, in the late prompt emission and in the early afterglow phases the most luminous EQTS regions are the ones closest to the boundary of the visible region. This transition in the emitting region may lead to specific observational signatures, i.e. an anomalous spectral evolution, in the rising part or at the peak of the prompt emission. We find as well an expression for the apparent radius of the EQTS in the sky, valid in both the fully radiative and the adiabatic regimes. Such considerations are essential for the theoretical interpretation of the prompt emission phase of GRBs.
\end{abstract}


\section{Introduction}

It is widely accepted that Gamma-Ray Burst (GRB) afterglows originate from the interaction of an ultrarelativistically expanding shell into the CircumBurst Medium (CBM). Differences exists on the detailed kinematics and dynamics of such a shell (see e.g. Refs.~\refcite{2005ApJ...633L..13B,2006RPPh...69.2259M} and Refs. therein).

Due to the ultrarelativistic velocity of the expanding shell (Lorentz gamma factor $\gamma \sim 10^2 - 10^3$), photons emitted at the same time in the laboratory frame (i.e. the one in which the center of the expanding shell is at rest) from the shell surface but at different angles from the line of sight do not reach the observer at the same arrival time. Therefore, if we were able to resolve spatially the GRB afterglows, we would not see the spherical surface of the shell. We would see instead the projection on the celestial sphere of the EQuiTemporal Surface (EQTS), defined as the surface locus of points which are source of radiation reaching the observer at the same arrival time (see e.g. Refs.~\refcite{1939AnAp....2..271C,1966Natur.211..468R,1998ApJ...494L..49S,1998ApJ...493L..31P,1999ApJ...513..679G,2001A&A...368..377B,2004ApJ...605L...1B,2005ApJ...620L..23B} and Refs. therein). The knowledge of the exact shape of the EQTSs is crucial, since any theoretical model must perform an integration over the EQTSs to compute any prediction for the observed quantities (see e.g. Refs.~\refcite{1999ApJ...511..852G,2004MNRAS.353L..35O,2004ApJ...605L...1B,2005ApJ...620L..23B,2005ApJ...618..413G,2006RPPh...69.2259M,2006ApJ...637..873H,2007ChJAA...7..397H} and Refs. therein).

One of the key problems is the determination of the angular size of the visible region of each EQTS, as well as the distribution of the luminosity over such a visible region. In the current literature it has been shown that in the latest afterglow phases the luminosity is maximum at the boundaries of the visible region and that the EQTS must then appear as expanding luminous ``rings'' (see e.g. Refs.~\refcite{1997ApJ...491L..19W,1998ApJ...494L..49S,1998ApJ...493L..31P,1999ApJ...513..679G,1999ApJ...527..236G,1998ApJ...497..288W,2003ApJ...585..899G,2003ApJ...593L..81G,2004ApJ...609L...1T,2008MNRAS.390L..46G} and Refs. therein). Such an analysis is applied only in the latest afterglow phases to interpret data from radio observations \cite{1997Natur.389..261F,1998ApJ...497..288W,2003ApJ...585..899G,2004ApJ...609L...1T,2005ApJ...618..413G,2005ApJ...622..986T,2007ApJ...664..411P} or gravitational microlensing \cite{2000ApJ...544L..11G,2001ApJ...561..178G,2001ApJ...561..703I,2001ApJ...551L..63G}. The shell dynamics is usually assumed to be fully adiabatic and to be described by a power-law $\gamma \propto r^{-3/2}$, following the Blandford-McKee self similar solution\cite{1976PhFl...19.1130B}, where $\gamma$ and $r$ are respectively the Lorentz gamma factor and the radius of the expanding shell. Such a power-law behavior has been extrapolated backward from the latest phases of the afterglow all the way to the prompt emission phase.

In Refs.~\refcite{2004ApJ...605L...1B,2005ApJ...620L..23B,2005ApJ...633L..13B} there have been presented the analytic solutions of the equations of motion for GRB afterglow, compared with the above mentioned approximate solutions, both in the fully radiative and adiabatic regimes, and the corresponding analytic expressions for the EQTSs. It has been shown that the approximate power-law regime can be asymptotically reached by the Lorentz gamma factor only in the latest afterglow phases, when $\gamma \sim 10$, and only if the initial Lorentz gamma factor $\gamma_\circ$ of the shell satisfies $\gamma_\circ > 10^2$ in the adiabatic case or $\gamma_\circ > 10^4$ in the radiative case. Therefore, in no way the approximate power-law solution can be used to describe the previous dynamical phases of the shell, which are the relevant ones for the prompt emission and for the early afterglow.

Starting from these premises, in this Paper we present the distribution of the extended afterglow luminosity over the visible region of a single EQTSs within the ``fireshell'' model for GRBs. Such a model uses the exact solutions of the fireshell equations of motion and assumes a fully radiative dynamics (see Refs.~\refcite{2001ApJ...555L.113R,2009AIPC.1132..199R} and Refs. therein for details). We recall that within the fireshell model the peak of the extended afterglow encompasses the prompt emission. We focus our analysis on the prompt emission and the early afterglow phases. Our approach is therefore complementary to the other ones in the current literature, which analyze only the latest afterglow phases, and it clearly leads to new results when applied to the prompt emission phase. For simplicity, we consider only the bolometric luminosity\cite{2002ApJ...581L..19R}, since during the prompt phase this is a good approximation of the one observed e.g. by BAT or GBM instruments\cite{2002ApJ...581L..19R,2004IJMPD..13..843R}. The analysis is separately performed over different selected EQTSs. The temporal evolution of the luminosity distribution over the EQTSs' visible region is presented. As a consequence of these results, we show the novel feature that at the beginning of the prompt emission the most luminous regions of the EQTSs are the ones closest to the line of sight. On the contrary, in the late prompt emission and in the early afterglow phases the most luminous EQTS regions are the ones closest to the boundary of the visible region. This transition in the emitting region may lead to specific observational signatures, i.e. an anomalous spectral evolution, in the rising part or at the peak of the prompt emission. We also present an analytic expression for the temporal evolution, measured in arrival time, of the apparent radius of the EQTSs in the sky. We finally discuss analogies and differences with other approaches in the current literature which assumes an adiabatic dynamics instead of a fully radiative one.

\section{The Equitemporal Surfaces (EQTS)}

For the case of a spherically symmetric fireshell considered in this Letter, the EQTSs are surfaces of revolution about the line of sight. The general expression for their profile, in the form $\vartheta = \vartheta(r)$, corresponding to an arrival time $t_a$ of the photons at the detector, can be obtained from (see e.g. Ref.~\refcite{2005ApJ...620L..23B}):
\begin{equation}
ct_a = ct\left(r\right) - r\cos \vartheta  + r^\star\, ,
\label{ta_g}
\end{equation}
where $r^\star$ is the initial size of the expanding fireshell, $\vartheta$ is the angle between the radial expansion velocity of a point on its surface and the line of sight, $t = t(r)$ is its equation of motion, expressed in the laboratory frame, and $c$ is the speed of light.

In the case of a fully radiative regime, the dynamics of the system is given by the following solution of the equations of motion (see e.g. Refs.~\refcite{1999PhR...314..575P,2005ApJ...620L..23B} and Refs. therein):
\begin{equation}
\gamma=\frac{1+\left(M_\mathrm{cbm}/M_B\right)\left(1+\gamma_\circ^{-1}\right)\left[1+\left(1/2\right)\left(M_\mathrm{cbm}/M_B\right)\right]}{\gamma_\circ^{-1}+\left(M_\mathrm{cbm}/M_B\right)\left(1+\gamma_\circ^{-1}\right)\left[1+\left(1/2\right)\left(M_\mathrm{cbm}/M_B\right)\right]}\, ,
\label{gamma_rad}
\end{equation}
where $\gamma$ is the Lorentz gamma factor of the fireshell, $M_\mathrm{cbm}$ is the amount of CBM mass swept up within the radius $r$ and $\gamma_\circ$ and $M_B$ are respectively the values of the Lorentz gamma factor and of the mass of the fireshell at the beginning of the extended afterglow phase. Correspondingly, the exact analytic expression for $t=t(r)$ is\cite{2005ApJ...620L..23B}:
\begin{equation}
\begin{split}
t\left(r\right) = & \tfrac{(M_B  - m_i^\circ)(r - r_\circ)}{2c\sqrt C } + \tfrac{{r_\circ \sqrt{3C}}}{{6 c m_i^\circ A^2 }} \left[\arctan \tfrac{2r- Ar_\circ}{Ar_\circ\sqrt 3 } - \arctan \tfrac{{2 - A}}{{A\sqrt 3 }}\right] \\[6pt] & + \tfrac{{r_\circ \sqrt C }}{{12cm_i^\circ A^2 }} \ln \left\{ {\tfrac{{\left[ {A + \left(r/r_\circ\right)} \right]^3 \left(A^3  + 1\right)}}{{\left[A^3  + \left( r/r_\circ \right)^3\right] \left( {A + 1} \right)^3}}} \right\} + t_\circ + \tfrac{m_i^\circ r_\circ }{8c\sqrt C }\left(\tfrac{r^4-r_\circ^4}{r_\circ^4}\right)\, ,
\end{split}
\label{analsol}
\end{equation}
where $A=\sqrt[3]{(M_B-m_i^\circ)/m_i^\circ}$, $C={M_B}^2(\gamma_\circ-1)/(\gamma_\circ +1)$, $t_\circ$ is the value of the time $t$ at the beginning of the extended afterglow phase and $m_i^\circ=(4/3)\pi m_p n_{\mathrm{cbm}} r_\circ^3$. Inserting Eq.(\ref{analsol}) into Eq.(\ref{ta_g}) we have the analytic expression for the EQTS in the fully radiative regime\cite{2005ApJ...620L..23B}:
\begin{equation}
\begin{split}
& \cos\vartheta = \tfrac{(M_B  - m_i^\circ)(r - r_\circ)}{2r\sqrt{C}} +\tfrac{{r_\circ \sqrt{3C} }}{{6rm_i^\circ A^2 }} \left[ \arctan \tfrac{2r- Ar_\circ}{Ar_\circ\sqrt 3 } - \arctan \tfrac{{2 - A}}{{A\sqrt{3} }}\right] \\[6pt]
&+ \tfrac{m_i^\circ r_\circ }{8r\sqrt{C}}\left(\tfrac{r^4-r_\circ^4}{r_\circ^4}\right)+\tfrac{{r_\circ \sqrt{C} }}{{12rm_i^\circ A^2 }} \ln \left\{ {\tfrac{{\left[ {A + \left(r/r_\circ\right)} \right]^3 \left(A^3  + 1\right)}}{{\left[A^3  + \left( r/r_\circ \right)^3\right] \left( {A + 1} \right)^3}}} \right\} +\tfrac{ct_\circ - ct_a + r^\star}{r}.
\end{split}
\label{eqts_g_dopo}
\end{equation}

Instead, the corresponding equations in the adiabatic regime are\cite{2005ApJ...620L..23B}:
\begin{equation}
\gamma^2=\frac{\gamma_\circ^2+2\gamma_\circ\left(M_\mathrm{cbm}/M_B\right)+\left(M_\mathrm{cbm}/M_B\right)^2}{1+2\gamma_\circ\left(M_\mathrm{cbm}/M_B\right)+\left(M_\mathrm{cbm}/M_B\right)^2}\, ,
\label{gamma_ad}
\end{equation}
\begin{equation}
t\left(r\right) = \left(\gamma_\circ-\tfrac{m_i^\circ}{M_B}\right)\tfrac{r-r_\circ}{c\sqrt{\gamma_\circ^2-1}} + \tfrac{m_i^\circ}{4M_Br_\circ^3}\left(\tfrac{r^4-r_\circ^4}{c\sqrt{\gamma_\circ^2-1}}\right) + t_\circ\, ,
\label{analsol_ad}
\end{equation}
\begin{equation}
\cos\vartheta = \tfrac{m_i^\circ}{4M_B\sqrt{\gamma_\circ^2-1}}\left(\tfrac{r^4-r_\circ^4}{r_\circ^3 r}\right) + \tfrac{ct_\circ - ct_a + r^\star}{r} - \tfrac{\gamma_\circ-\left(m_i^\circ/M_B\right)}{\sqrt{\gamma_\circ^2-1}}\left(\tfrac{r_\circ-r}{r}\right)\, .
\label{eqts_g_dopo_ad}
\end{equation}

\begin{figure}
\includegraphics[width=\hsize,clip]{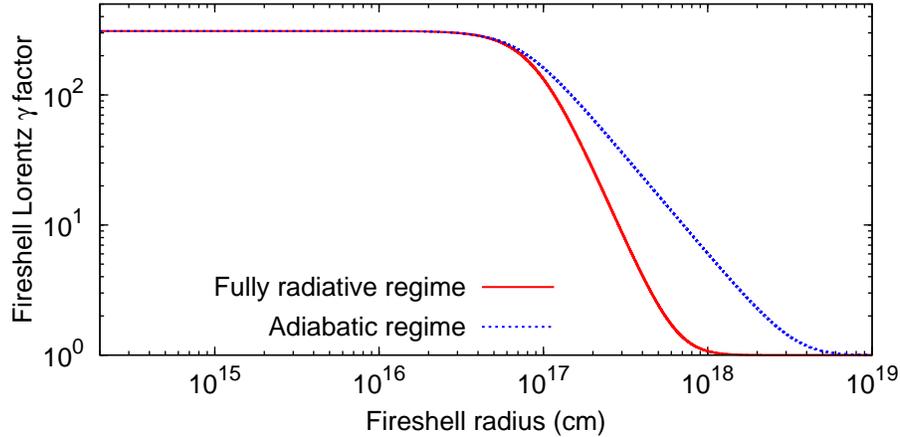}
\caption{The Fireshell Lorentz gamma factors in the fully radiative regime (red line), given by Eq.(\ref{gamma_rad}), and in the adiabatic regime (blue line), given by Eq.(\ref{gamma_ad}).}
\label{gamma_rad_ad}
\end{figure}

A comparison between the Lorentz gamma factors in the two regimes is presented in Fig.~\ref{gamma_rad_ad}. Here and in the following we assume the same initial conditions as in Ref.~\refcite{2005ApJ...620L..23B}, namely $\gamma_\circ = 310.131$, $r_\circ = 1.943 \times 10^{14}$ cm, $t_\circ = 6.481 \times 10^{3}$ s, $r^\star = 2.354 \times 10^8$ cm, $n_{cbm} = 1.0$ particles/cm$^3$, $M_B = 1.61 \times 10^{30}$ g. For simplicity, and since we are interested in the overall behavior of the luminosity distribution, we assume a constant CBM density, neglecting the inhomogeneities which are responsible of the temporal variability of the prompt emission \cite{2002ApJ...581L..19R}.

\section{The extended afterglow luminosity distribution over the EQTS}\label{bld}

\begin{figure*}
\centering
\includegraphics[width=0.495\hsize,clip]{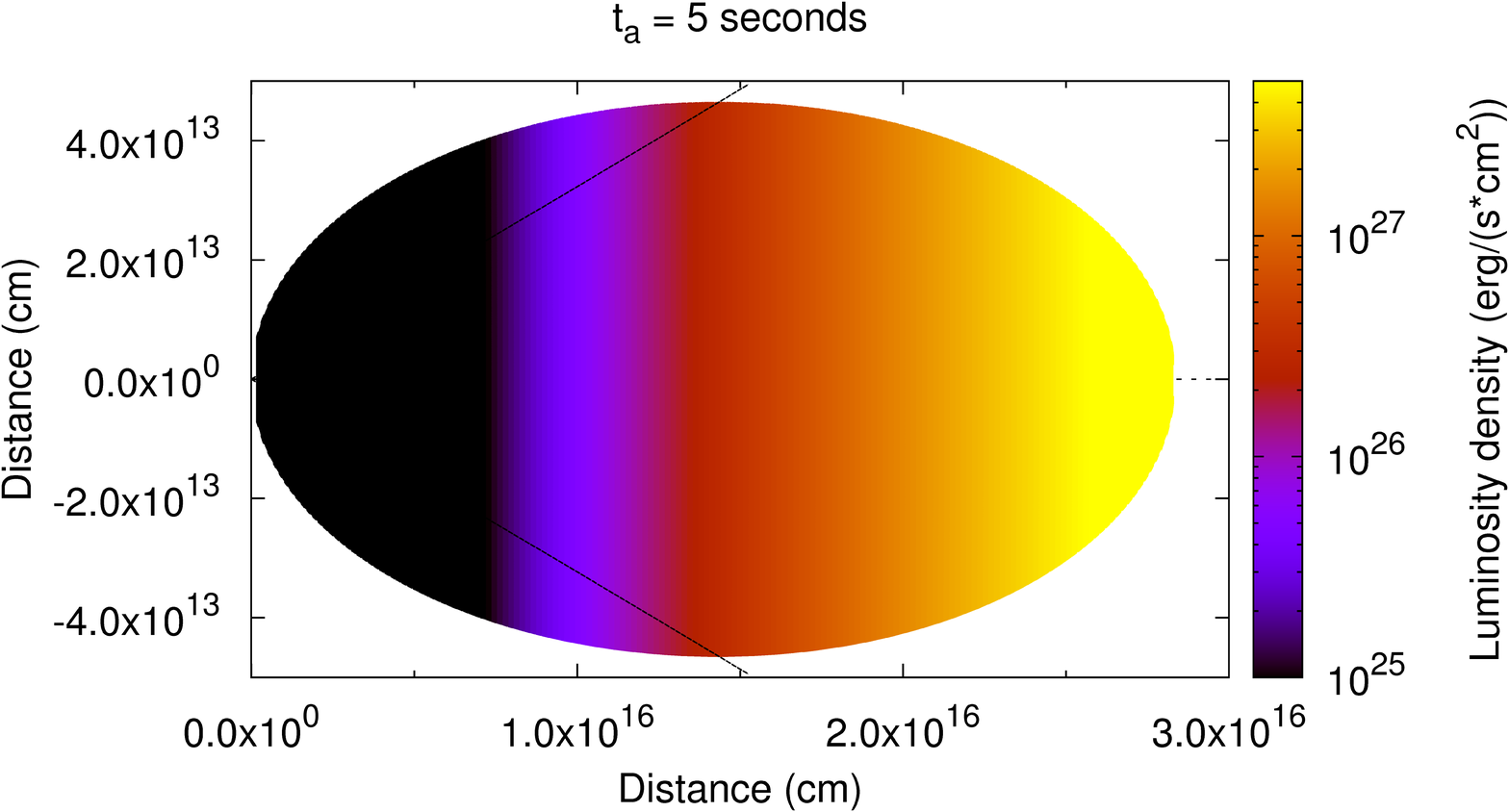}\hfill
\includegraphics[width=0.495\hsize,clip]{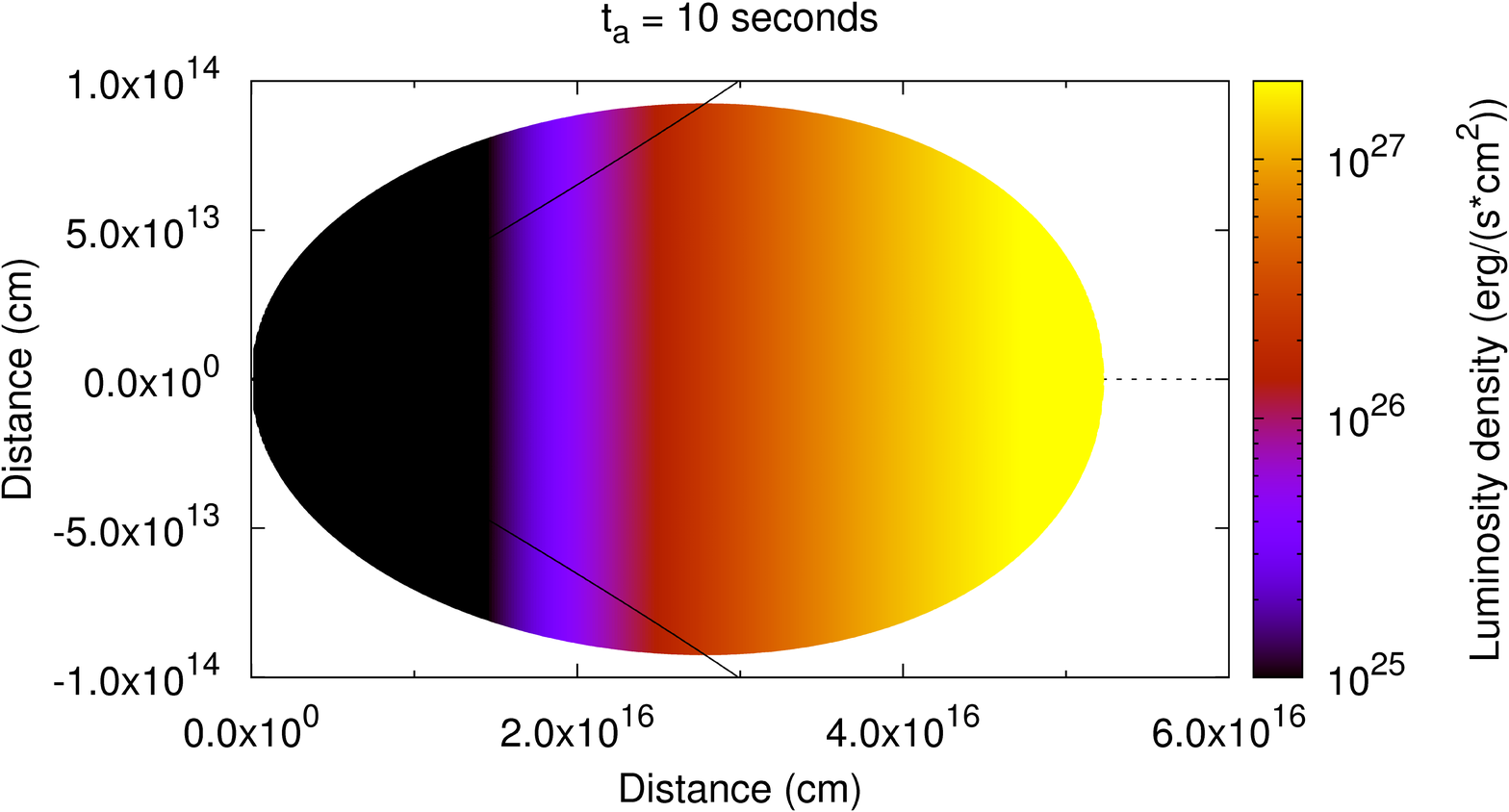}\\
\includegraphics[width=0.495\hsize,clip]{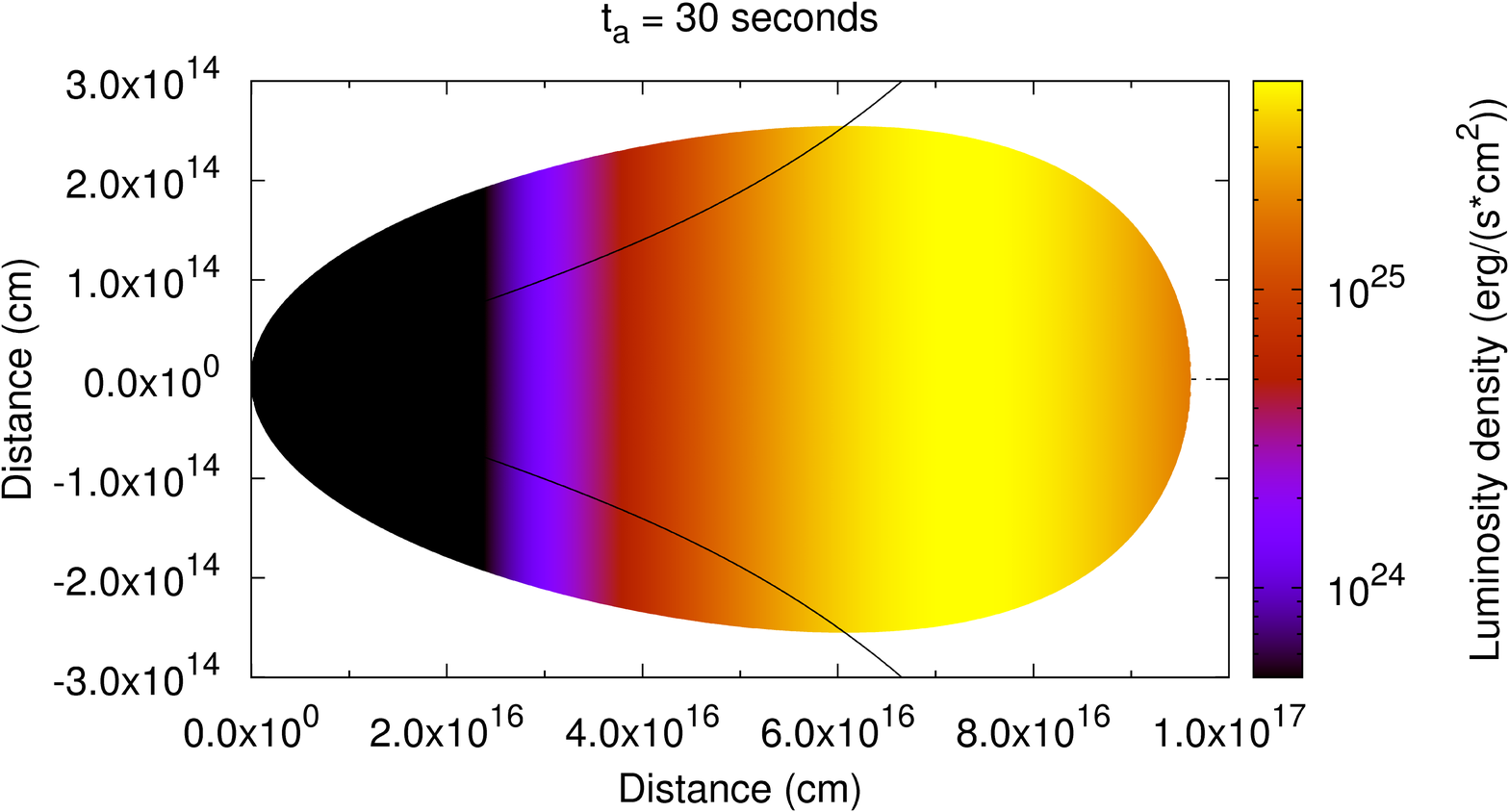}\hfill
\includegraphics[width=0.495\hsize,clip]{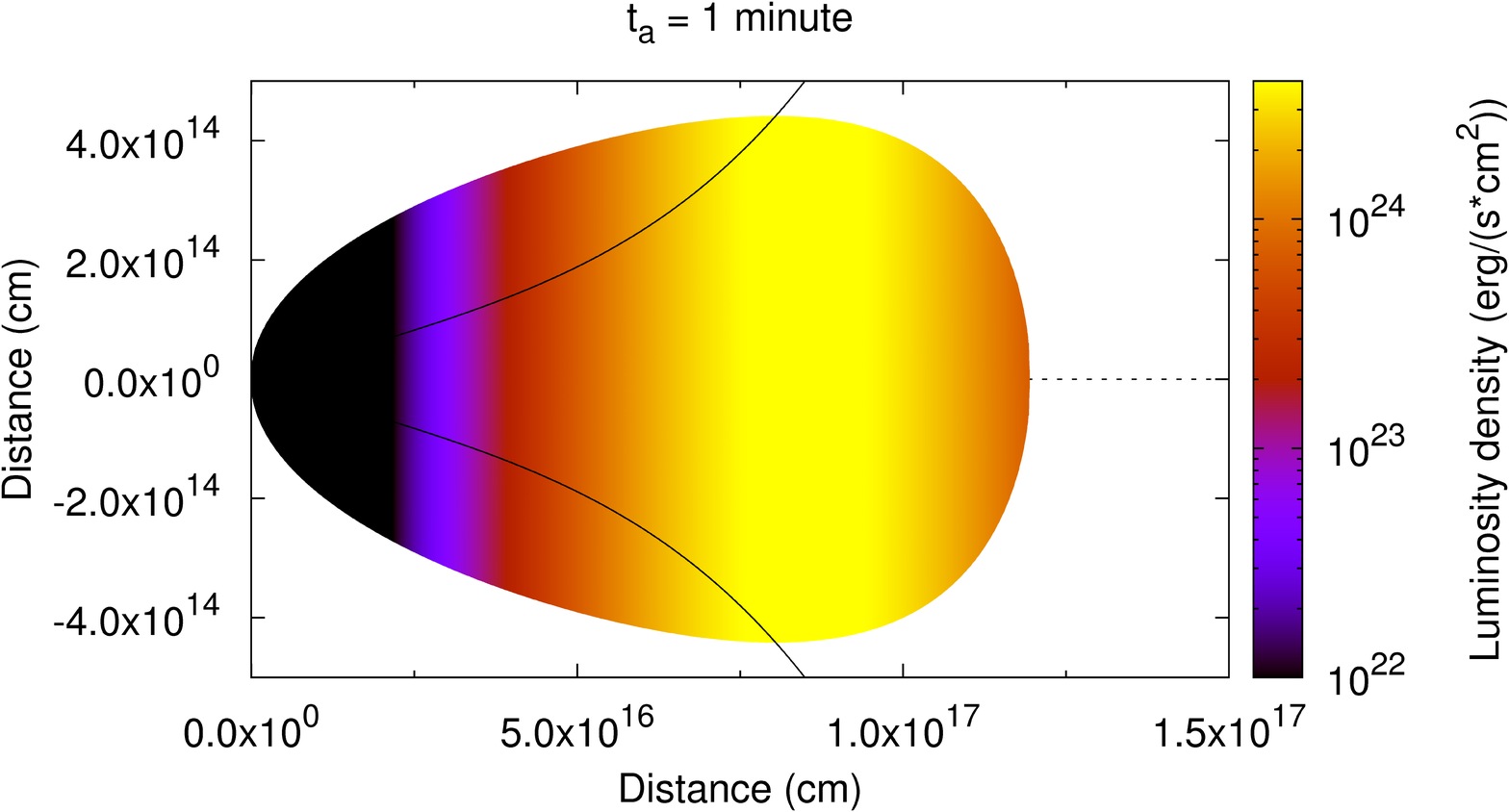}\\
\includegraphics[width=0.495\hsize,clip]{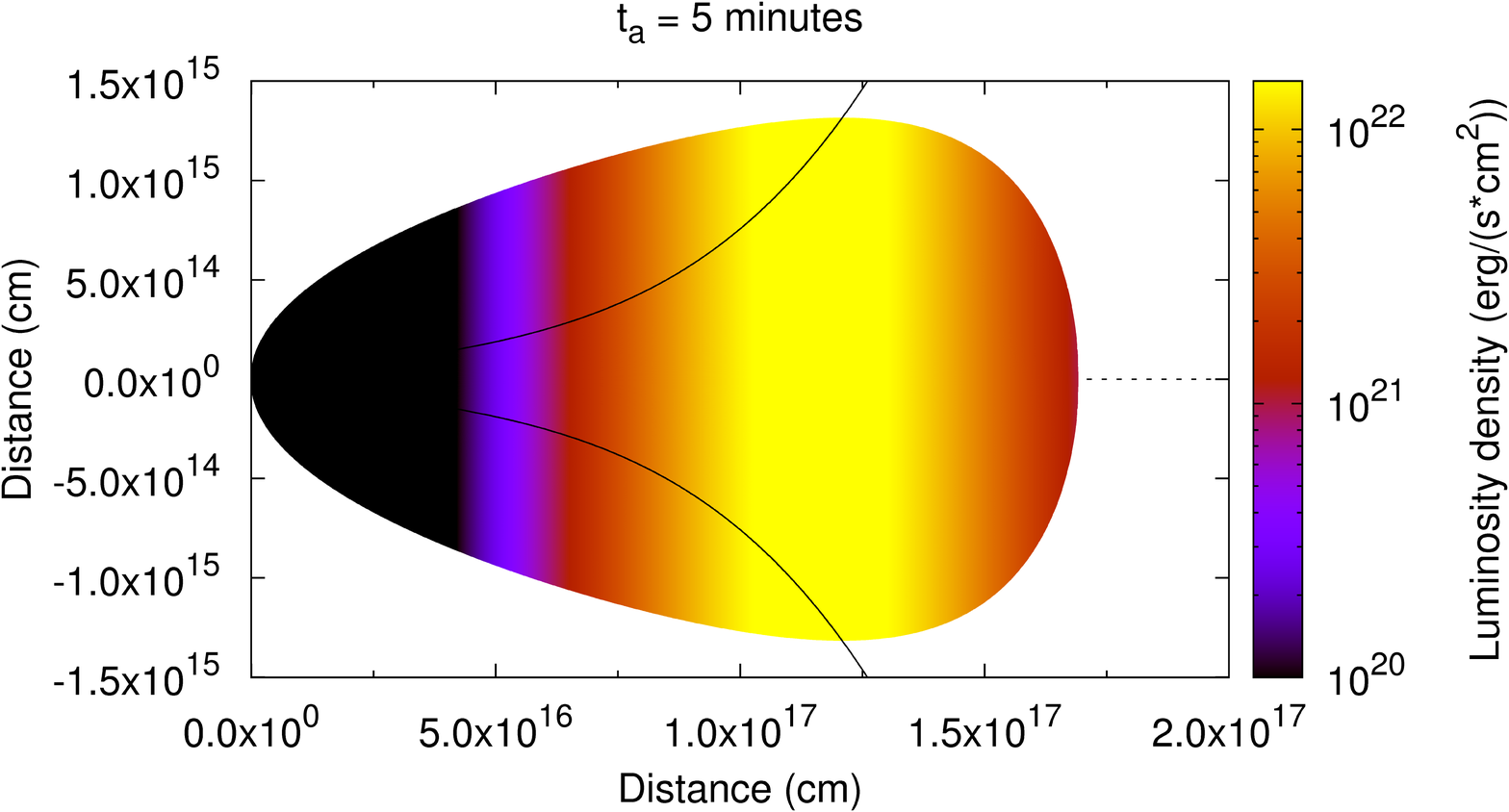}\hfill
\includegraphics[width=0.495\hsize,clip]{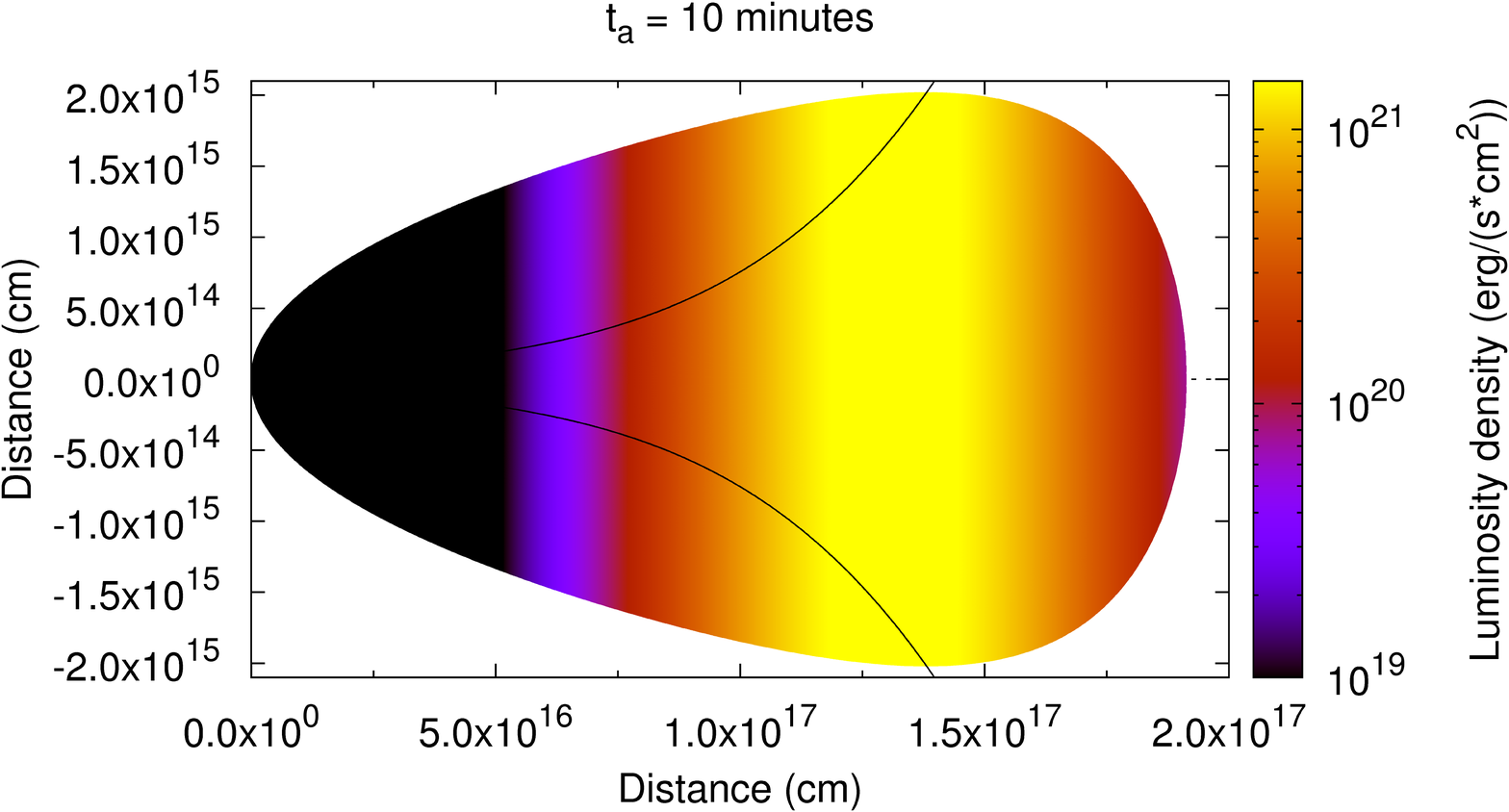}\\
\includegraphics[width=0.495\hsize,clip]{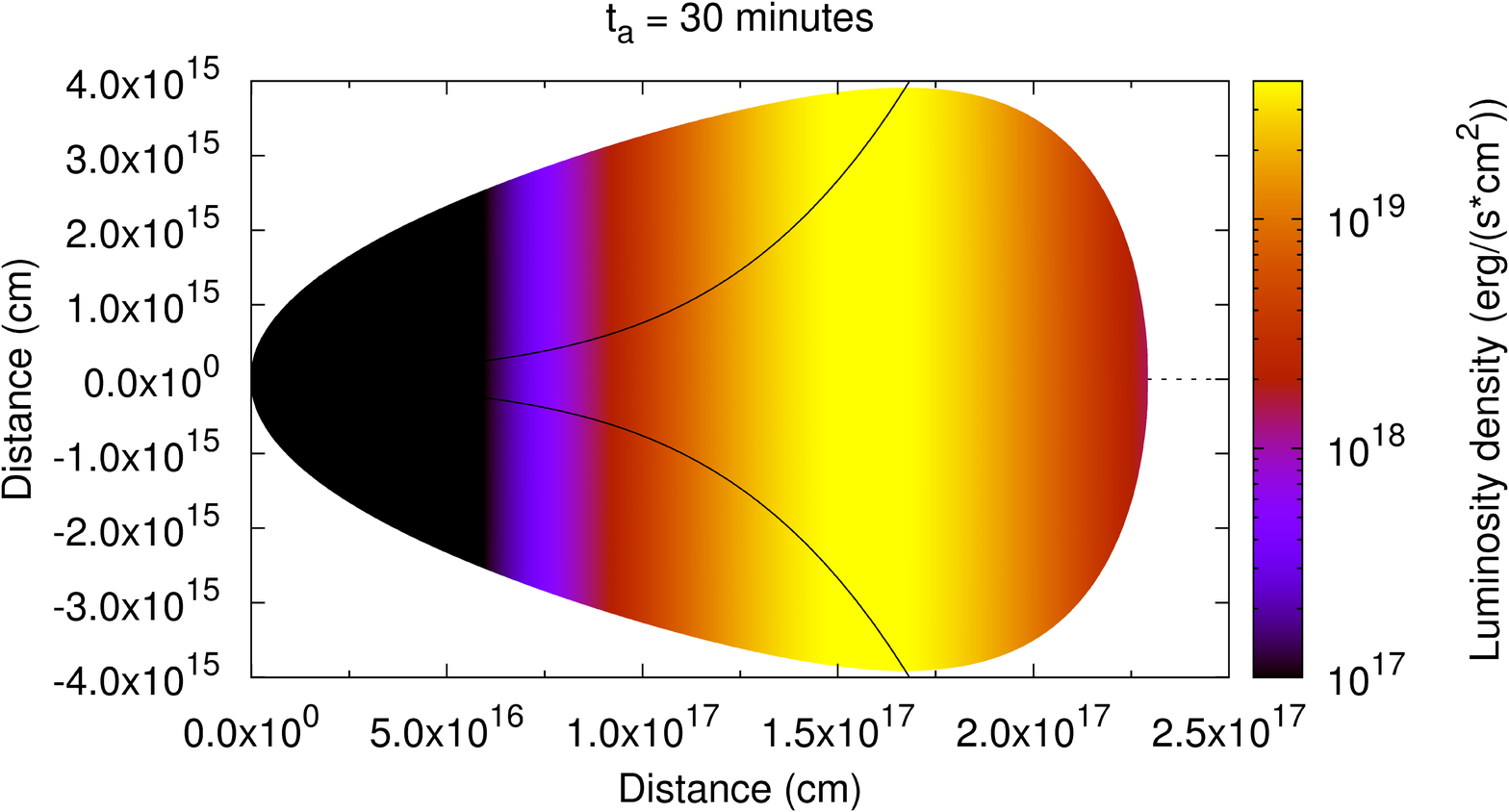}\hfill
\includegraphics[width=0.495\hsize,clip]{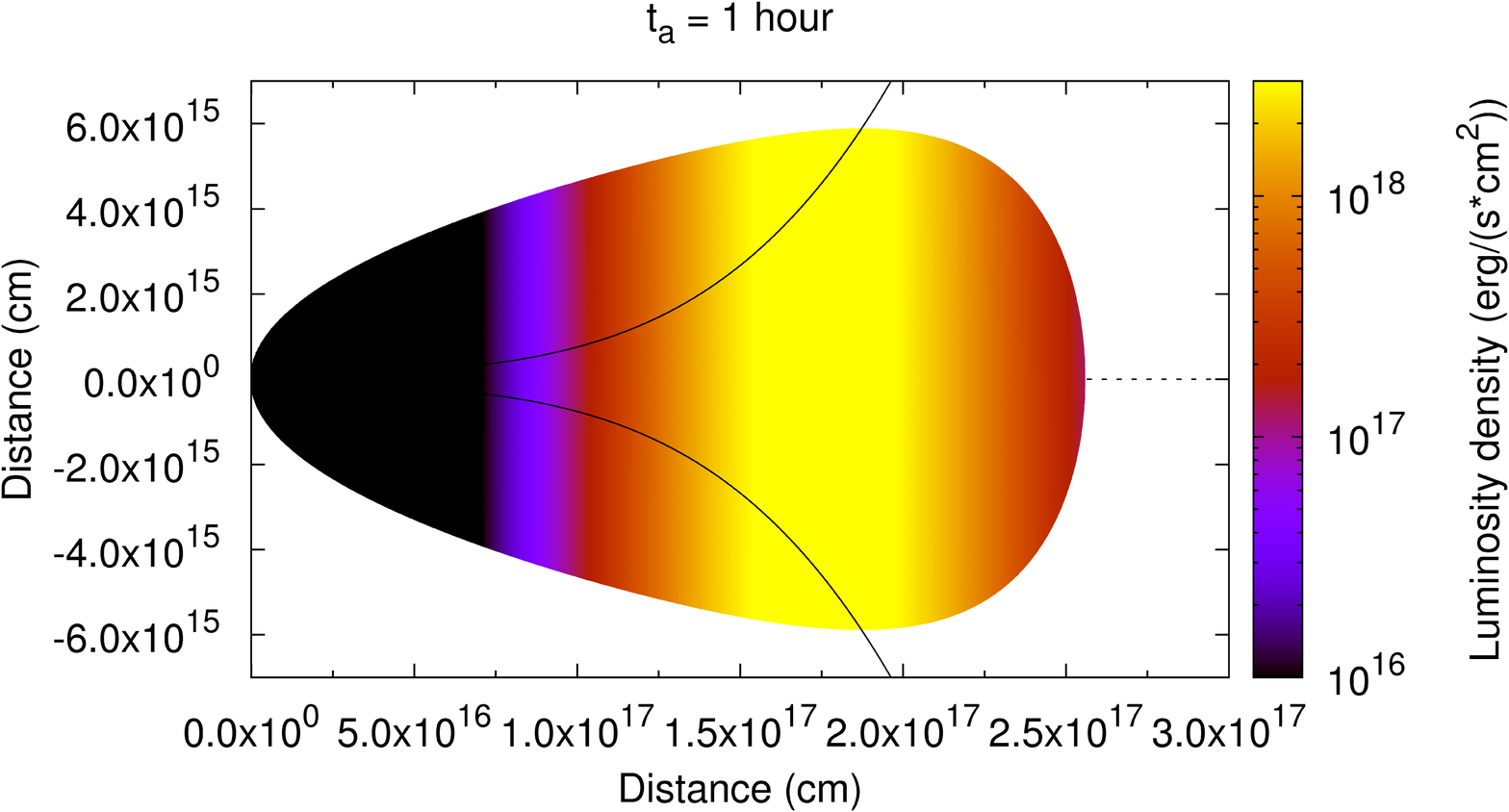}
\caption{We plot the luminosity density $\mathcal{L}$ given in Eq.(\ref{fluxarr}) over 8 different EQTSs, corresponding to 8 different $t_a$ values ranging from the prompt emission ($5$ seconds) to the early ($1$ hour) afterglow phases (see title above each plot). Each EQTS is represented as its projection on a plane containing the line of sight, which coincides with the X axis of each plot and which is represented by a dashed gray line. The observer is far away along the X axis. The luminosity density distribution over each EQTS is represented by a color gradient, with the highest values corresponding to the lightest colors (see the key on the left of each plot). The scales of the different axes and of the color gradient are different among the different plots, since it was not possible to choose a single scale suitable for all of them. The black curves represent the condition $\cos\vartheta = v/c$, see Eq.(\ref{vis}). The boundaries of the visible regions are therefore defined by the intersections of such lines with the EQTS external profiles, which coincide with the points where the EQTS external profiles show an horizontal tangent in the plots.}
\label{EQTS_multi}
\end{figure*}

Within the fireshell model, the GRB extended afterglow bolometric luminosity in an arrival time $dt_a$ and per unit solid angle $d\Omega$ is given by (details in Refs.~\refcite{2002ApJ...581L..19R,2009AIPC.1132..199R} and Refs. therein):
\begin{equation}
\frac{dE_\gamma}{dt_a d\Omega } \equiv \int_{EQTS} \mathcal{L}(r,\vartheta,\varphi;t_a) d\Sigma = \int_{EQTS} \frac{\Delta \varepsilon \cos \vartheta v dt}{4 \pi \Lambda^{4} dt_a} d \Sigma,
\label{fluxarr}
\end{equation}
where $\Delta \varepsilon$ is the energy density released in the interaction of the ABM pulse with the CBM measured in the comoving frame, $\Lambda=\gamma(1-(v/c)\cos\vartheta)$ is the Doppler factor, $d\Sigma$ is the surface element of the EQTS at arrival time $t_a$ on which the integration is performed, and it has been assumed the fully radiative condition. We are here not considering the cosmological redshift of the source, which is constant during the GRB explosion and therefore it cannot affect the results of the present analysis. We recall that in our case such a bolometric luminosity is a good approximation of the one observed in the prompt emission and in the early afterglow by e.g. the BAT or GBM instruments\cite{2002ApJ...581L..19R,2004IJMPD..13..843R}.

We are now going to show how this luminosity is distributed over the EQTSs, i.e. we are going to plot over selected EQTSs the luminosity density $\mathcal{L}(r,\vartheta,\varphi;t_a)$. The results are represented in Fig. \ref{EQTS_multi}. We chose eight different EQTSs, corresponding to arrival time values ranging from the prompt emission ($5$ seconds) to the early ($1$ hour) afterglow phases. For each EQTS we represent also the boundaries of the visible region due to relativistic collimation, defined by the condition (see e.g. Refs.~\refcite{2006ApJ...644L.105B} and Refs. therein):
\begin{equation}
\cos\vartheta \ge v/c\, .
\label{vis}
\end{equation}
We obtain that, at the beginning ($t_a = 5$ seconds), when $\gamma$ is approximately constant, the most luminous regions of the EQTS are the ones along the line of sight. However, as $\gamma$ starts to drop ($t_a \gtrsim 30$ seconds), the most luminous regions of the EQTSs become the ones closest to the boundary of the visible region. This transition in the emitting region may lead to specific observational signatures, i.e. an anomalous spectral evolution, in the rising part or at the peak of the prompt emission.

\section{The EQTS apparent radius in the sky}

\begin{figure}
\includegraphics[width=\hsize,clip]{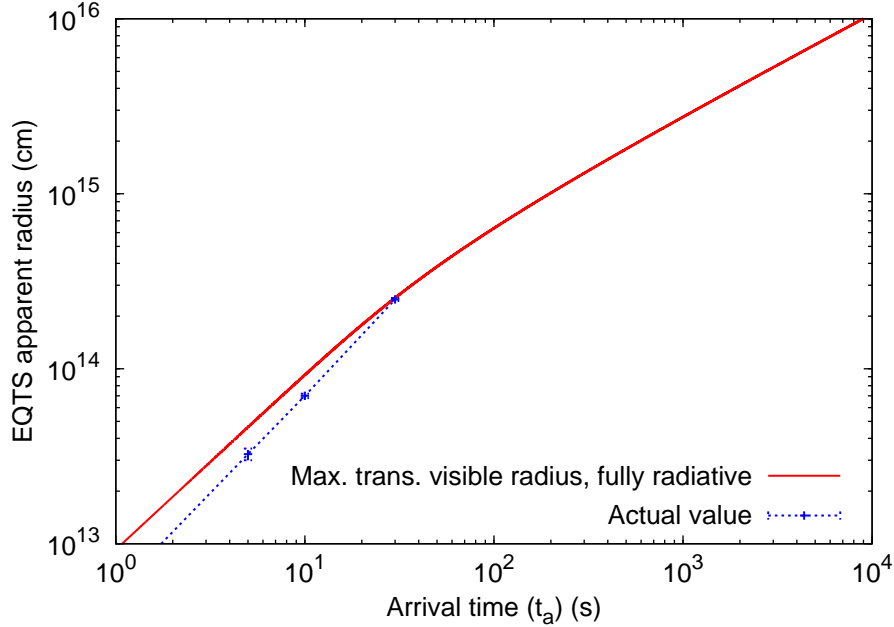}
\caption{The EQTS apparent radius in the sky as a function of the arrival time $t_a$ in the fully radiative regime. The red line represents the maximum transverse EQTS visible radius $r_\bot$ following Eq.(\ref{rbotfin}). The three blue points represents the actual EQTS apparent radius in the sky in the three cases $t_a = 5.0$ s, $t_a = 10.0$ s and $t_a = 30.0$ s in which, as shown in Fig.~\ref{EQTS_multi}, it is smaller than $r_\bot$. The blue line is a linear interpolation of such points.}
\label{rmaxvsta_rad}
\end{figure}

From sec.~\ref{bld} we obtain that within the fireshell model the EQTSs of GRB extended afterglows should appear in the sky as point-like sources at the beginning of the prompt emission but evolving after a few seconds into expanding luminous ``rings'', with an apparent radius evolving in time and always equal to the maximum transverse EQTS visible radius $r_\bot$ which can be obtained from Eqs.(\ref{ta_g},\ref{vis}):
\begin{equation}
\left\{\begin{array}{l}
r_\bot = r \sin\vartheta\\
ct_a = ct\left(r\right) - r\cos \vartheta  + r^\star\\
\cos\vartheta = v/c
\end{array}\right.\, .
\label{rbot1}
\end{equation}
where $t = t(r)$ is given by Eq.(\ref{analsol}). With a small algebra we get:
\begin{equation}
\left\{\begin{array}{l}
r_\bot = r/\gamma(r)\\
t_a = t(r) - (r/c)\sqrt{1-\gamma(r)^{-2}}  + (r^\star/c)
\end{array}\right.\, ,
\label{rbotfin}
\end{equation}
where $\gamma\equiv\gamma\left(r\right)$ is given by Eq.(\ref{gamma_rad}) and $t\equiv t \left(r\right)$ is given by Eq.(\ref{analsol}), since we assumed the fully radiative condition. Eq.(\ref{rbotfin}) defines parametrically the evolution of $r_\bot \equiv r_\bot \left(t_a\right)$, i.e. the evolution of the maximum transverse EQTS visible radius as a function of the arrival time. In Fig.~\ref{EQTS_multi} we saw that such $r_\bot$ coincides with the actual value of the EQTS apparent radius in the sky only for $t_a \gtrsim 30$ s, since for $t_a \lesssim 30$ s the most luminous EQTS regions are the ones closest to the line of sight (see the first three plots in Fig.~\ref{EQTS_multi}). Therefore, in Fig.~\ref{rmaxvsta_rad} we plot $r_\bot$ given by Eq.(\ref{rbotfin}) in the fully radiative regime together with the actual values of the EQTS apparent radius in the sky taken from Fig.~\ref{EQTS_multi} in the three cases in which they are different. It is clear that, during the early phases of the prompt emission, even the ``exact solution'' given by Eq.(\ref{rbotfin}) can be considered only an upper limit to the actual EQTS apparent radius in the sky.

\begin{figure}
\includegraphics[width=\hsize,clip]{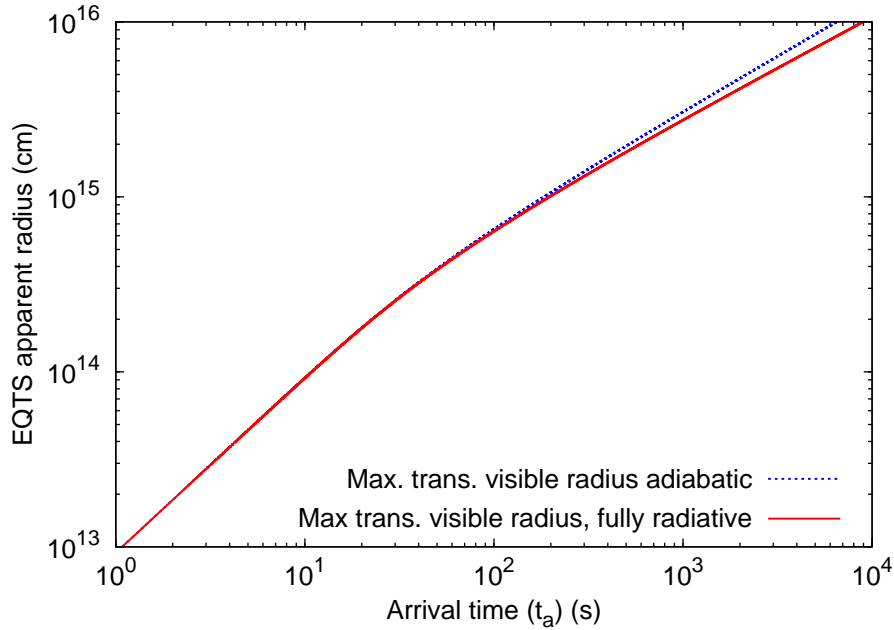}
\caption{Comparison between the maximum transverse EQTS visible radius $r_\bot$ computed with Eq.(\ref{rbotfin}) in the adiabatic (blue line) and in the fully radiative (red line) cases respectively.}
\label{rmaxvsta_rad_ad}
\end{figure}

In the current literature (see e.g. Refs.~\refcite{1998ApJ...494L..49S,1998ApJ...497..288W,1999ApJ...513..679G,1999ApJ...527..236G,2000ApJ...544L..11G,2001ApJ...551L..63G,2001ApJ...561..178G,2003ApJ...585..899G,2003ApJ...593L..81G,2004ApJ...609L...1T,2004MNRAS.353L..35O,2005ApJ...618..413G,2008MNRAS.390L..46G}) there are no analogous treatments, since it is always assumed an adiabatic dynamics instead of a fully radiative one and only the latest afterglow phases are addressed. It is usually assumed the Blandford-McKee self similar solution\cite{1976PhFl...19.1130B} for the adiabatic dynamics $\gamma \propto r^{-3/2}$. A critical analysis of the applicability to GRBs of this approximate dynamics, instead of the exact solutions in Eqs.(\ref{gamma_ad},\ref{analsol_ad},\ref{eqts_g_dopo_ad}), has been presented in Ref.~\refcite{2005ApJ...633L..13B}, as recalled in the introduction. The most widely applied formula for the EQTS apparent radius in the above mentioned current literature is the one proposed by Sari\cite{1998ApJ...494L..49S}:
\begin{equation}
r_\bot = 3.91 \times 10^{16} \left(E_{52}/n_1\right)^{1/8}\left[T_{days}/(1+z)\right]^{5/8} \, \mathrm{cm},
\label{gps99}
\end{equation}
where $E_{52}$ is the initial energy of the shell in units of $10^{52}$ ergs, $n_1$ is the CBM density in units of $1$ particle/cm$^3$, $T_{days}$ is the arrival time at the detector of the radiation measured in days, and $z$ is the source cosmological redshift. Waxman et al.\cite{1998ApJ...497..288W} however derived a numerical factor of $3.66 \times 10^{16}$ instead of $3.91\times 10^{16}$. Eq.(\ref{gps99}) cannot be compared directly with Eq.(\ref{rbotfin}), since they assume two different dynamical regimes. Therefore, we first plot in Fig.~\ref{rmaxvsta_rad_ad} the exact solution for $r_\bot$ given by Eq.(\ref{rbotfin}) both in the fully radiative case, using Eqs.(\ref{gamma_rad},\ref{analsol}), and in the adiabatic one, using Eqs.(\ref{gamma_ad},\ref{analsol_ad}). We see that they almost coincide during the prompt emission while they start to diverge in the early afterglow phases, following the behavior of the corresponding Lorentz gamma factors (see Fig.~\ref{gamma_rad_ad} and details in Ref.~\refcite{2005ApJ...633L..13B}).

\begin{figure}
\includegraphics[width=\hsize,clip]{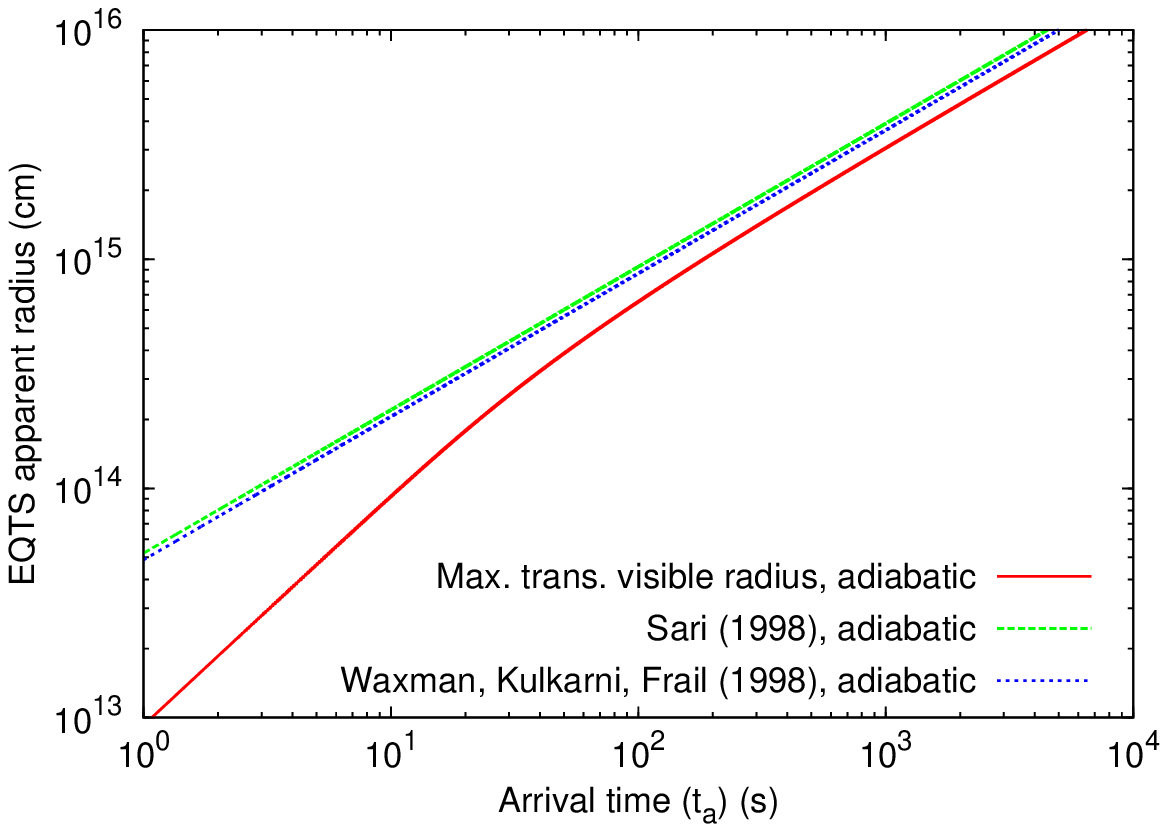}
\caption{The EQTS apparent radius in the sky as a function of the arrival time $t_a$ in the adiabatic regime. The red line represents the maximum transverse EQTS visible radius $r_\bot$ following Eq.(\ref{rbotfin}). The green line represents Eq.(\ref{gps99}) proposed by Sari. The blue line represents the corresponding modification in the numerical factor proposed by Waxman et al. These last two lines are almost coincident on the scale of this plot.}
\label{rmaxvsta_ad}
\end{figure}

Now, in Fig.~\ref{rmaxvsta_ad} and for the same initial conditions assumed in previous figures, we plot the maximum transverse EQTS visible radius $r_\bot$ given by Eq.(\ref{rbotfin}) in the adiabatic case, using Eqs.(\ref{gamma_ad},\ref{analsol_ad}), together with Eq.(\ref{gps99}), proposed in Ref.~\refcite{1998ApJ...494L..49S}, and with the corresponding modification in the numerical factor proposed in Ref.~\refcite{1998ApJ...497..288W}. From such a comparison, we can see that the approximate regime $r_\bot \propto t_a^{5/8}$ overestimates the exact solution for $r_\bot$ during the prompt emission and the early afterglow phases. It is asymptotically reached only in the latest afterglow phases ($t_a \gtrsim 10^3$ s), in the very small region in which the approximate power-law dynamics starts to be applicable (see details in Ref.~\refcite{2005ApJ...633L..13B}). However, there is still a small discrepancy in the normalization: the constant numerical factor in front of Eq.(\ref{gps99}) should be $\sim 3.1\times 10^{16}$ instead of $3.91 \times 10^{16}$ or $3.66\times 10^{16}$ to reproduce the behavior of the exact solution for large $t_a$. Moreover, we must emphasize that, in analogy with what we obtained in the fully radiative case (see Fig.~\ref{rmaxvsta_rad}), in the early phases of the prompt emission, when the fireshell Lorentz $\gamma$ factor is almost constant, the maximum transverse EQTS visible radius $r_\bot$ given by Eq.(\ref{rbotfin}) can only be considered an upper limit to the actual value of the EQTS apparent radius in the sky. In such phases the approximation implied by Eq.(\ref{gps99}) can therefore be even worse that what represented in Fig.~\ref{rmaxvsta_ad}.

\section{Conclusions}

Within the fireshell model, using the exact analytic expressions for the fireshell equations of motion and for the corresponding EQTSs in the fully radiative condition, we analyzed the temporal evolution of the distribution of the extended afterglow luminosity over the EQTS during the prompt emission and the early afterglow phases. We find that, at the beginning of the prompt emission ($t_a = 5$ seconds), when $\gamma$ is approximately constant, the most luminous regions of the EQTS are the ones along the line of sight. As $\gamma$ starts to drop ($t_a \gtrsim 30$ seconds), the most luminous regions of the EQTSs become the ones closest to the boundary of the visible region. This transition in the emitting region may lead to specific observational signatures, i.e. an anomalous spectral evolution, in the rising part or at the peak of the prompt emission. The EQTSs of GRB extended afterglows should therefore appear in the sky as point-like sources at the beginning of the prompt emission but evolving after a few seconds into expanding luminous ``rings'', with an apparent radius evolving in time and always equal to the maximum transverse EQTS visible radius $r_\bot$.

We also derive an exact analytic expression for $r_\bot$, both in the fully radiative and in the adiabatic conditions, and we compared it with the approximate formulas commonly used in the current literature in the adiabatic case. We found that these last ones can not be applied in the prompt emission nor in the early afterglow phases. Even when the asymptotic regime is reached ($t_a \gtrsim 10^3$ s), it is necessary a correction to the numerical factor in front of the expression given in Eq.(\ref{gps99}) which should be $\sim 3.1\times 10^{16}$ instead of $3.91 \times 10^{16}$ or $3.66\times 10^{16}$.

\end{document}